# Observation of Custodial Chiral Symmetry in Memristive Topological Insulators

Wenhao Li,[1,2,3,4] Junyao Wu,[1,2,3,4] Ning Han,[5] Rui Zhao,[1,2,3,4] Fujia Chen,[1,2,3,4] Yuang Pan,[1,2,3,4] Yudong Ren,[1,2,3,4] Hongsheng Chen,[1,2,3,4] Zhen Gao,[6,*] Ce Shang,[7,†] and Yihao Yang[1,2,3,4,‡]

[1] *Interdisciplinary Center for Quantum Information State Key Laboratory of Extreme Photonics and Instrumentation ZJU-Hangzhou Global Scientific and Technological Innovation Center, Zhejiang University, Hangzhou 310000, China*

[2] *International Joint Innovation Center The Electromagnetics Academy at Zhejiang University. Zhejiang University, Haining 314400, China*

[3] *Key Lab. of Advanced Micro/Nano Electronic Devices Smart Systems of Zhejiang Jinhua Institute of Zhejiang University, Zhejiang University, Jinhua 321000, China*

[4] *Shaoxing Institute of Zhejiang University, Zhejiang University, Shaoxing 312000, China*

[5] *College of Optical and Electronic Technology, China Jiliang University, Hangzhou 310018, China*

[6] *State Key Laboratory of Optical Fiber and Cable Manufacturing Technology, Department of Electronic and Electrical Engineering, Guangdong Key Laboratory of Integrated Optoelectronics Intellisense, Southern University of Science and Technology, Shenzhen 518055, China and*

[7] *Aerospace Information Research Institute, Chinese Academy of Sciences, Beijing 100094, China*

The concept of custodial symmetry, a residual symmetry that protects physical observables from large quantum corrections, has been a cornerstone of high-energy physics, but its experimental observation has remained unexplored. Building on recent theoretical work [Phys. Rev. Lett. 128, 097701 (2022)], we report the first experimental observation of classical analog of custodial chiral symmetry in a memristive Su-Schrieffer-Heeger (SSH) circuit. We provide direct experimental evidence for custodial symmetry through the measurement of the correction to the Lagrangian. This Lagrangian correction, which mimics a mass term in field theory, vanishes smoothly as the perturbation is reduced. We also demonstrate that topological edge states in the memristive SSH circuit remain localized at the boundary, protected by custodial chiral symmetry. This work opens new avenues for emulating field-theoretic symmetries and nonlinear dynamics in memristive platforms.

*Introduction.*—The Standard Model of quantum field theory (QFT) has been remarkably successful in describing the strong and electroweak interactions. However, it remains incomplete, lacking a unified framework and a fundamental explanation for the origin of particle masses. To address these issues, various symmetries are often considered [1,2]. Among them, custodial symmetry, a residual global SU(2) symmetry emerging from the Higgs sector, plays a key role in stabilizing the electroweak sector. It protects the mass ratio between the *W* and *Z* bosons by suppressing large radiative corrections, ensuring that quantum corrections to certain observables vanish in the symmetric limit [3–6].

In QFT, when a system possesses custodial symmetry, quantum corrections to a parameter are proportional to the parameter itself [5,7], typically manifested as a symmetry-breaking term in the system's Lagrangian ($\mathcal{L}$), as shown in Fig. 1 (see Supplementary Material for additional discussion [8]). The concept has inspired theoretical efforts to realize custodial symmetry in engineered systems [5], although no experimental observation has yet been achieved.

Topology has emerged as a powerful framework for understanding symmetry-protected states of matter, with wide-ranging applications in electronics [10,11], photonics [12–14], phononics [15,16], and electrical circuits [17–19]. For example, in the Su-Schrieffer-Heeger (SSH) model, which represents a one-dimensional (1D) topological insulator, the in-gap topological edge states are protected by the chiral symmetry which refers to the existence of a unitary operator that anticommutes with the SSH Hamiltonian, ensuring that the energy spectrum is symmetric about zero and enabling the definition of a topological winding number [5,20]. Since the topological phases fundamentally rely on certain symmetries, understanding what happens when these symmetries are explicitly broken becomes crucial. Motivated by this, a recent theoretical work has proposed introducing custodial symmetry mechanisms into topological systems to suppress the effects of symmetry-breaking corrections [5]. However, such ideas remain untested in experiments.

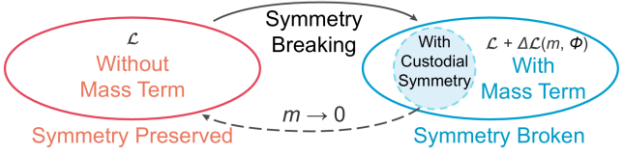

FIG. 1 Illustration of the custodial symmetry. The symmetry preserved states are encircled with an orange line, whose Lagrangian is denoted as $\mathcal{L}$. By inducing the symmetry broken term, the Lagrangian of the symmetry broken states (encircled by a solid blue line) is corrected by a mass-dependent term $\Delta\mathcal{L}$, where $\tilde{\mathcal{L}} = \mathcal{L} + \Delta\mathcal{L}(m,\phi)$. When the system holds custodial symmetry (encircled by a dashed blue line), the correction is suppressed in the limit that $m \to 0$.

In this work, we experimentally observe custodial symmetry, specifically custodial chiral symmetry, in an SSH circuit augmented with memristors. Memristors, often regarded as a fundamental circuit element alongside resistors, capacitors, and inductors, are widely utilized as nonlinear components capable of generating complex dynamical behavior [21–24]. The introduction of memristors explicitly breaks the chiral symmetry inherent in the original SSH model; however, the resulting memristive SSH circuit exhibits a residual global symmetry, which we identify as custodial chiral symmetry. In experiments, we first explore the chaotic behavior of a two-resonator circuit with memristive coupling, confirming the existence of divergent, quasi-periodic, and damped states. We then construct memristive SSH circuits and directly measure the correction to the Lagrangian under various memristor configurations, which shows that the Lagrangian correction mimics a mass term in field theory and vanishes smoothly as the perturbation is reduced. We also demonstrate that the topological edge states in the memristive SSH model remain localized at the boundary, protected by the custodial chiral symmetry.

*Results.*—To characterize the nonlinearity of memristor, we consider a temporal coupled mode system connecting two resonators with an effective coupling strength $\kappa_M$ (See details in Supplementary Materials [8]) [Fig. 2(a)]. Here, we construct a coupled two-resonator inductor–capacitor (*LC*) circuit whose coupling capacitor is replaced by a memristor, whose concept was proposed by Chua in 1971 [Fig. 2(b)] [21]. As illustrated in Fig. 2(c), the memristor couples the charge ($q$) and flux ($\varphi$) via the constitutive relation $d\varphi = Mdq$ [21,25–27]. We apply the input and output voltages at the ports of two *LC* resonant circuits, and the input and output resistances of the applied voltages are denoted by $R$.

Due to the nonlinear characteristics, memristors have been, and are being, exploited as fundamental blocks for the definition of new nonlinear circuits able to show complex behavior [22,28]. Following the postulation of the ideal memristor, the description of the memristance $M$ in terms of the voltage integration across the memristor, as given by [29]:

$$M = \alpha[\int(V_1 - V_2)dt + \beta] \quad (1)$$

where $V_1$ and $V_2$ denote the voltages at each node of the memristor, and $\alpha$, $\beta$ are the constants (see details in the Supplementary Material [8]). In this configuration, a voltage is applied at the input node ($V_1$), and the voltage at the output node ($V_2$) is observed.

By tuning the internal constants $\beta$, we can switch between distinct dynamical regimes. By solving the resulting nonlinear differential equations, we analytically obtain the temporal signals for the system. As illustrated in Figs. 2(d-f), when $\alpha = 1$, the Lissajous curves of $V_2$-$V_1$ for $\beta = 0.7, 0.8$, and $0.9$ exhibit distinct voltage states, that is, $V_2$ transitions through divergent, quasi-periodic, and damped states, respectively [30]. Experimentally, instead of using integrated memristive devices, we implement the memristor via an active equivalent circuit. This strategy avoids device-to-device variability and limited endurance in real memristive devices, and provides a reproducible state-dependent resistance [31]. In the circuit realization, the constant $\alpha$ corresponds to the amplifier characteristics, while $\beta$ is controlled by the direct current offset (see details in the Supplementary Material [8]). The measured temporal signals are in excellent agreement with the calculated results, as shown in Figs. 2(g-i). Notably, when $\beta = 0.8$, both the calculated and measured Lissajous curves reveal strange attractors, which serve as evidence of chaotic dynamic behavior [23]. To further analyze the chaotic flows, we computed the Lyapunov exponents of the measured $V_2$ for $\beta = 0.7$, 0.8, and 0.9, respectively. The corresponding values of the Lyapunov exponents are 0.219, 0.024, and −0.198, respectively, providing further confirmation of the system's chaotic dynamics [32]. These results confirm that the memristor provides a nonlinear interaction, whose strength can be modulated in situ. Such nonlinearity not only enables the engineering of chaotic flows [33] but also mimics the behavior of nonequilibrium systems. Such nonlinearity of the memristor provides a versatile foundation for exploring symmetry breaking and topological phenomena, as discussed in the following sections.

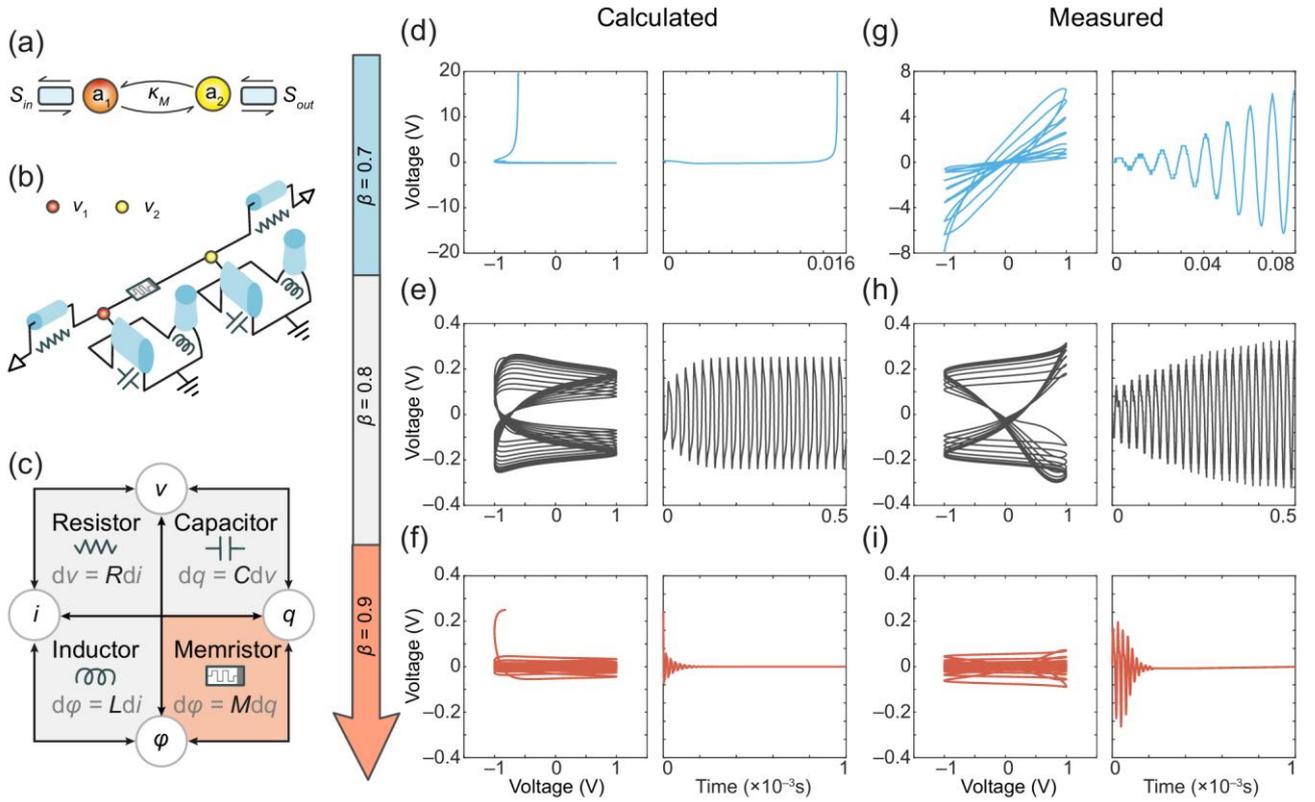

FIG. 2. Two-resonator circuit with memristive coupling. (a) Memristive temporal coupling model. (b) Illustration of the coupling circuit. Red and yellow dots mark nodes $V_1$ and $V_2$, respectively. Memristor is implemented by an active equivalent circuit. (c) Conceptual symmetries of resistor, capacitor, inductor, and memristor. (d-f) Calculated and (g-i) measured Lissajous curves (left panel) of $V_2$-$V_1$ and voltage waveforms (right panel) of $V_2$ when $\beta = 0.7$, 0.8, and 0.9. Largest Lyapunov exponents are 0.219, 0.024, and −0.198, respectively.

Having established that the memristor introduces a tunable nonlinear response, we consider the 1D SSH circuit with memory as a prototypical example, where memristors are integrated in parallel with the capacitors [Figs. 3(a-b)]. Here, the memristors are also implemented by active equivalent circuits, and provide a reproducible for probing symmetry-breaking effects [31]; In principle, the same SSH architecture could be realized on-chip using integrated memristive devices. In the SSH circuit, the system consists of an alternating series of two types of capacitors, $C_1$ and $C_2$, and inductors $L$. The topological bandgap and edge states are determined by the ratio $r = C_2 / C_1$, with $r < 1$ yielding a topological bandgap [17,20,34,35]. For a representative configuration ($C_1 = 2.2$ μF, $C_2 = 1.0$ μF, $L = 100$ μH), the measured band structure shows a clear topological bandgap (grey region) matching well with the calculated results, which spans from 8.0 to 11.5 kHz, as shown in Fig. 3(c). Under open boundary conditions, the eigenstates of the system can be derived by solving the eigenvalue problem $\mathbf{I} = \mathbf{GV}$, where $\mathbf{G}$ is the conductance matrix, $\mathbf{I}$ and $\mathbf{V}$ are currents and voltages at each node, respectively (see Supplementary Material for details [8,17]). The midgap edge state appears as an exponentially localized mode at the odd-numbered nodes, while even-numbered nodes exhibit zero voltage (see Supplementary Material for detailed temporal voltage profiles [8]).

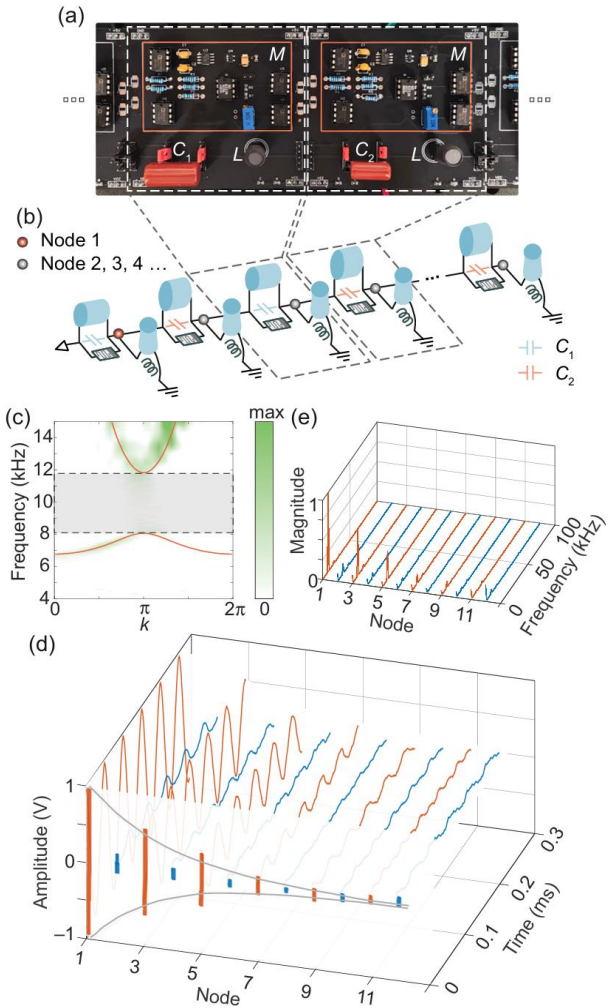

FIG. 3. SSH circuit with memristors. (a) Photograph of the memristive SSH circuit. (b) Illustration of the SSH circuit with memristors in parallel with the capacitors. The node 1 and the nodes 2, 3, 4 … are marked by red and grey dots, respectively. (c) Calculated (orange lines) and measured (colormap) band structure of the SSH circuit, when the chiral symmetry is preserved. The bandgap is marked by the gray area. (d-e) The measured temporal voltage profiles (d) and Fourier spectra (e) of voltage at each node with memristors, when $v_1 = 8.9$ kHz. The gray sheet in (d) is an exponentially attenuated envelope of unperturbed edge modes, and the amplitude projection is shown in the front.

The SSH model is a prototypical symmetry-protected topological insulator, meaning that the topological invariant is associated with chiral symmetry [34]. According to topological theory, this edge state is robust against perturbations that preserve chiral symmetry. However, when memristors are introduced into the SSH circuit, the original chiral symmetry is explicitly broken by the nonlinearity. As a result, the voltage waveforms of the nonlinear SSH circuit exhibit prominent components at the even nodes, and overtop the exponentially attenuated envelope (grey sheet), along with additional frequency components [Figs. 3(d-e)], in contrast to the clean edge mode in the unperturbed circuit within the bandgap [Fig. 3(c)] (see Supplementary Material for detailed temporal voltage profiles [8]). This indicates the emergence of a new behavior due to the perturbations introduced by the memristors and the breakdown of chiral symmetry. These results confirm that memristors introduce a symmetry-breaking perturbation that alters the topological response.

To further quantify this effect, we employ the Lagrangian of the circuit elements to more precisely describe the memristor-induced symmetry breaking in the SSH circuit [36]. From the action principle, standard variational calculus yields the Euler-Lagrange (EL) equations of motion. When memristors are added in parallel with $C_1$ and $C_2$, the system gains an additional dissipative term. The corresponding Lagrangian with correction ($\tilde{\mathcal{L}}$) can be expressed as a perturbative correction to the original one ($\mathcal{L}$),

$$\tilde{\mathcal{L}} = \mathcal{L} + \frac{1}{2}\sum_n \frac{1}{i\omega M_n}\left(\dot{\varphi}(n) - \dot{\varphi}(n-1)\right)^2 \quad (2)$$

where $n$ represents the number of nodes, $\omega$ denotes the frequency, dot denotes the time derivative, and $\varphi(n)$ represents the flux of the $n$-th node (see Supplementary Material for details [8]). The increase in the Lagrangian corresponding to the symmetry-breaking correction in our model. It is noteworthy that the Lagrangian for the SSH circuit with memristors can be viewed as a correction to the original expression, where the correction term is proportional to $1/M_n$, with $M_n$ representing the memristance at node $n$. By increasing the magnitude of the memristances, the violation of the original chiral symmetry in the SSH circuit can be suppressed.

This functional form emulates the structure of a "mass term" in QFT, where symmetry-breaking corrections are proportional to the symmetry-breaking parameter itself. As such, the SSH circuit exhibits a classical analog of custodial symmetry: the correction vanishes smoothly as the memristor-induced perturbation is turned off, then preserving the topological state. This is precisely the definition of custodial symmetry. To further illustrate the impact of memristors, we consider a more intuitive model for memristive elements that captures the main features of the experimentally realized devices, since the memristors share the same dimension as resistors [5,25]:

$$R_M = R_{on}x + R_{off}(1-x) \quad (3)$$

where $R_M$ is the memory resistance, $R_{on}$ and $R_{off}$ are its limits, and $x \in [0, 1]$ is the internal state variable. Since $R_{on}$ corresponds to the low-resistance state, it plays a more crucial role in the memory resistance. For simplicity, we focus on memristors with varying values of $R_{on}$ and examine their nonlinear characteristics. It is worth noting that the resistance of memristors depends on the voltage on each node. Here, we employ the $R_{on}$ of the first memristor to represent the overall trend.

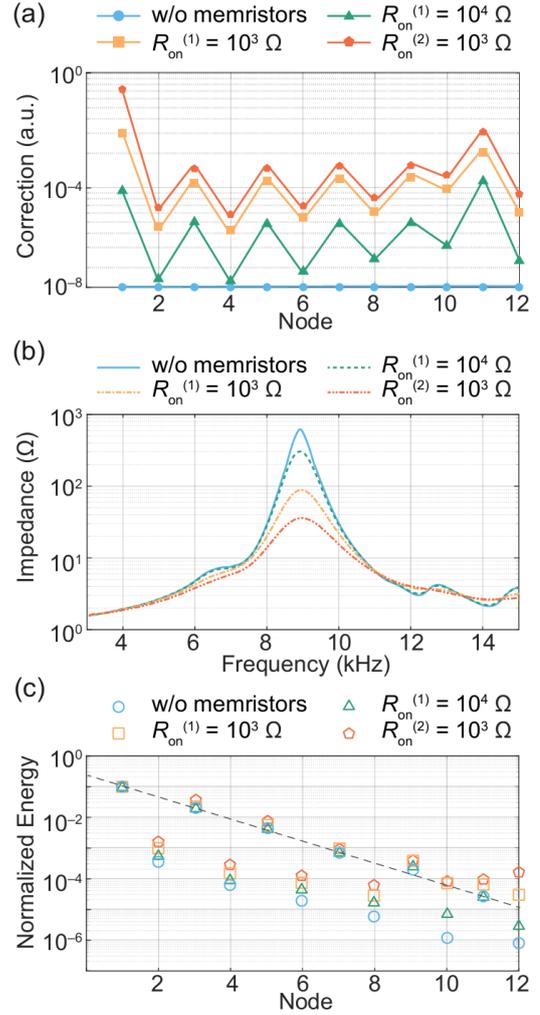

FIG. 4. Custodial chiral symmetry of SSH circuit with memristors. (a) Measured correction of Lagrangian without and with memristors for different values of the low memristance state, $R_{on}$. (b) Impedance between node 1 and the ground as a function of frequency of the circuit in Fig. 3(a), without and with memristors for different values of the low memristance state, $R_{on}$. (c) Normalized energy of voltage at each node of the same circuits in (b).

We design two types of memristors, labeled with superscripts corresponding to their respective $R_{on}$ values. Notably, the type 2 memristors exhibit stronger nonlinearity (see Supplementary Material for details of memristor design [8]). To quantitatively characterize the effect of symmetry breaking, we use the node-resolved Lagrangian correction $\Delta\mathcal{L}$ defined in Eq. (2). For each memristors configuration, we first measure the complex steady-state voltage $V_n(\omega)$ at all nodes and then evaluate, at each node $n$, the additional contribution

$\Delta \mathcal{L}_n$ obtained by inserting the measured voltages into Eq. (2) and subtracting unperturbed SSH Lagrangian. Since the time derivation of flux is equal to the voltage, the correction scales as $\Delta \mathcal{L}_n \propto |v(n)-v(n-1)|^2 / M_n$. We can see the resulting spatial profiles of $\Delta \mathcal{L}_n$ for different $R_{on}$ values and types of memristors, as shown in Fig. 4(a) (see more details in Supplementary Materials [8]). Node 0 is connected to ground and excluded from the analysis. The results show that increasing the memristance leads to a clear suppression of the correction, providing direct experimental evidence that the mass-like correction vanishes in the high-resistance limit, and consistent with the custodial symmetry mechanism.

To correlate this behavior with the topological properties of the circuit, we measure two key observables. First, the topological state of the SSH circuit is characterized by a peak in the impedance between Node 1 and ground as a function of frequency [Fig. 4(b)]. Second, the spatial localization of the edge state is confirmed by the energy distribution across the nodes [Fig. 4(c)]. As shown in Fig. 4(b), the impedance peak within the midgap diminishes with decreasing $R_{on}$ and stronger nonlinearity. A similar trend is observed in the distribution of the edge states [Fig. 4(c)], demonstrating the influence of symmetry-breaking perturbations on the topological states under custodial chiral symmetry. From these observations, we conclude that the custodial symmetry of the SSH circuit protects the topological edge state from large corrections. The violation of this symmetry is suppressed with a larger mass term, leading to enhanced stability of the topological edge states [3–5,7,37,38].

*Discussion.*—In summary, we have experimentally observed a classical analog of custodial symmetry in a topolectrical SSH circuit by introducing memristor-induced symmetry-breaking perturbations. To emulate the custodial symmetry, the associated Lagrangian correction acting as a mass-like term that vanishes in the high-resistance limit of memristors. Remarkably, these devices also give rise to controlled chaotic dynamics in a two-resonator circuit, further confirming the strong nonlinearity of the memristor. Both experimental and theoretical results demonstrate that this mechanism preserves topological edge states under perturbation. Our findings suggest that topolectrical memristive circuits offer a promising platform for studying both custodial symmetry and nonlinear dynamical phenomena in classical systems, and that the same circuit architecture can, in principle, be implemented on chip with integrated memristive devices, paving the way toward classical realizations of field-theoretic symmetry mechanisms.

*Acknowledgments*—The work was sponsored by the Key Research and Development Program of the Ministry of Science and Technology under Grants 2022YFA1404900 (Y.Y.), 2022YFA1404902 (H.C.), No.2022YFA1404704 (H.C.), and 2022YFA1405200 (Y.Y.), the National Natural Science Foundation of China (NNSFC) under Grants No. 62175215 (Y.Y.), No. 61975176 (H.C.), No. 62361166627 (Z.G.), and No. 62375118 (Z.G.), the Key Research and Development Program of Zhejiang Province under Grant No.2022C01036 (H.C.), the Fundamental Research Funds for the Central Universities (2021FZZX001-19) (Y.Y.), the Excellent Young Scientists Fund Program (Overseas) of China (Y.Y.), the Chinese Academy of Sciences project No. E4BA270100 (C.S.), No. E4Z127010F (C.S.), No. E4Z6270100 (C.S.), and No. E53327020D (C.S.), the Guangdong Basic and Applied Basic Research Foundation No. 2024A1515012770, Shenzhen Science and Technology Innovation Commission No. 202308073000209, and High-level Special Funds No. G03034K004.

*Data availability*—All data are available from the corresponding authors upon reasonable request.

‡ yangyihao@zju.edu.cn
∗ gaoz@sustech.edu.cn
† shangce@aircas.ac.cn